\documentclass[a4paper,11pt]{article}
\pdfoutput=1 
\usepackage{labelschanged}
\usepackage{jheppub} 

\usepackage[T1]{fontenc} 
\usepackage{xspace}
\usepackage{array}
\usepackage{multirow}
\usepackage{soul}
\usepackage{color}

\def\mathswitch#1{\relax\ifmmode#1\else$#1$\fi}
\def\mathswitchr#1{\relax\ifmmode{\mathrm{#1}}\else$\mathrm{#1}$\fi}

\newcommand{\as}{\hat\alpha_{\mathrm s}}

\newcommand{\msbar}{\ensuremath{\overline{\text{MS}}}\xspace}

\title{\bf Hadronic effects in M\o ller scattering at NNLO}

\author[a]{Jens~Erler,}
\author[a]{Rodolfo~Ferro-Hern\'andez}
\author[b]{and Ayres~Freitas}

\affiliation[a]{Institut für Kernphysik, Johannes Gutenberg-Universität Mainz, Mainz, Germany. }
\affiliation[b]{Department of Physics and Astronomy, University of Pittsburgh, Pittsburgh, PA 15260, 
USA.}

\emailAdd{erler@uni-mainz.de}
\emailAdd{afreitas@pitt.edu}
\emailAdd{rferrohe@uni-mainz.de}

\abstract{Two-loop electroweak corrections to polarized M\o ller scattering are studied in two different 
schemes at low energies. We find the finite $Q^2$ corrections to be well under control. 
The hadronic and perturbative QCD corrections to the $\gamma Z$ two-point function are incorporated 
through the weak mixing angle at low energies, which introduce an error of $0.08\times10^{-3}$ in the 
weak charge of the electron $Q^e_W$. 
Furthermore, by studying the scheme dependence, we obtain an estimate of the current perturbative 
electroweak uncertainty, $\delta Q^e_W\approx0.23\times10^{-3}$, which  is five times smaller than the 
precision estimated for the MOLLER experiment ($\delta Q_W^e=1.1\times10^{-3}$). 
Future work is possible to reduce the theory error further.}

\begin{document} 
\maketitle
\flushbottom
\section{Introduction}
\label{intro}

Parity-violating electron scattering is a powerful tool to test the Standard Model (SM) 
and probe for physics beyond the SM. Polarized M\o ller scattering,
\emph{viz.}\ the scattering of a polarized electron beam on electrons in a fixed target, 
offers the opportunity for high-precision measurements of the left-right asymmetry,
\begin{align}
A_{\rm LR} = \frac{d\sigma_{\rm L} - d\sigma_{\rm R}}{d\sigma_{\rm L} +
d\sigma_{\rm R}},
\end{align}
where the subscript L (R) refers to the left- (right)-handed polarization of the incident
electron beam. 
The MOLLER experiment \cite{Benesch:2014bas} currently under development at Jefferson Lab  
aims to determine $A_{\rm LR}$ with a relative precision of 2.4\%, which is an improvement 
by a factor of about five compared to the SLAC E158 result~\cite{Anthony:2005pm}.

This level of precision necessitates the inclusion of radiative corrections in the analysis. 
The theoretical prediction for $A_{\rm LR}$ can be written as~\cite{Derman:1979zc}
\begin{equation}
A_{\rm LR} = \frac{G_\mu Q^2}{\sqrt{2}\pi\alpha}\,\frac{1-y}{1+y^4+(1-y)^4}
(1-4\sin^2\theta_W + \Delta Q_W^e),
\label{eq:alr}
\end{equation}
where $G_\mu$ is the Fermi constant, $y=Q^2/s$, and $s$ and $Q^2$ are the (squares of the)
center-of-mass energy and the momentum transfer between the two electrons, respectively. 
The quantity $\Delta Q_W^e$ denotes the radiative corrections to the so-called weak charge of the 
electron $Q_W^e$, {\em i.e.}, the expression in parentheses in Eq.~(\ref{eq:alr}).

The next-to-leading order (NLO) corrections were found to be sizeable, reducing
the tree-level prediction for $A_{\rm LR}$ by approximately\footnote{This unusually 
large relative correction does not signal the breakdown of perturbation theory, 
but it is due to the fact that the leading-order (LO) contribution is accidentally 
small since $1-4\sin^2\theta_W \ll 1$.} 40\%~\cite{Czarnecki:1995fw}. 
This implies that higher-order corrections need to be considered to
match the anticipated precision of the MOLLER experiment. Recently, the electroweak
next-to-next-to-leading order (NNLO) from diagrams with closed fermion loops were
obtained~\cite{Du:2021zkj}, and they were found to have a
moderate impact of 1.3\% relative to the LO asymmetry.

At the one-loop level~\cite{DePorcel:1995nh,Czarnecki:1995fw}, the numerically
dominant contribution stems from the $\gamma Z$ mixing self-energy,
which contains logarithmically enhanced terms $\propto \ln(m_f^2/m_Z^2)$, where
$f$ is any electrically charged fermion in the SM. An additional complication
arises from the fact that the light quark contribution to the $\gamma Z$
self-energy is not well-defined, since $Q^2 < \Lambda^2_{\rm QCD}$ and thus
non-perturbative hadronization effects become important.

It is well known that these large logarithms (and therefore also the leading
hadronic uncertainty) can be absorbed by expressing the one-loop result in
terms of the \msbar weak mixing angle\footnote{We use carets to denote quantities in the \msbar scheme.} 
at the scale zero, $\hat{s}^2(0)$, rather
than at the weak scale~\cite{Czarnecki:1998xc,Erler:2004in,Erler:2017knj}.
This approach also offers the opportunity to resum higher-order QCD corrections
by using renormalization-group (RG) techniques for the computation of the
running of $\hat{s}^2(Q^2)$ in the perturbative $Q^2$-regime~\cite{Erler:2004in,Erler:2017knj}. 
It is desirable to apply this strategy also at higher orders by
recasting the recent electroweak NNLO result in terms of $\hat{s}^2(0)$. To
accomplish this, one must expand the shift from RG running, $\Delta\hat{s}^2
\equiv \hat{s}^2(0)-\hat{s}^2(m_Z)$, in fixed orders of perturbation theory
and adjust the explicit one- and two-loop contributions to $\Delta\hat{s}^2$ in
the NNLO result of Ref.~\cite{Du:2021zkj}.

In this way, one arrives at the most accurate description of M\o ller scattering
in the limit $Q^2 \to 0$. It should be noted, however, that there can be 
corrections for realistic $Q^2 \neq 0$. For diagrams with only massive $W$ and
$Z$ bosons, the $Q^2$ dependence is suppressed by powers of $Q^2/m_{W,Z}^2$ and
thus completely negligible. However, this hierarchy of scales does not apply to
the $\gamma\gamma$ and $\gamma Z$ self-energies and thus the $Q^2$ dependence 
cannot be ignored here.
Ref.~\cite{Czarnecki:1995fw} observed that at NLO there are large cancellations among the residual
$Q^2 \neq 0$ loop corrections, and the remainder was estimated to be numerically small. 
Since it is not clear whether similar cancellations occur at NNLO,
a more detailed investigation is needed.

This paper addresses both of the issues mentioned above: 
(a) the use of the low-scale \msbar weak mixing angle $\hat{s}^2(0)$ within the
electroweak NNLO correction to M\o ller scattering, 
and (b) the investigation of non-zero $Q^2$ effects. 
Section~\ref{rg} describes the two-loop expansion of the RG running of $\hat{s}^2$. 
In particular, the extraction of the fixed-order shift $\Delta\hat{s}^2$ from the RG study 
of Ref.~\cite{Erler:2017knj} will be discussed in detail. 
In Section~\ref{oneloop}, we show how this replacement works for the one-loop result of 
Ref.~\cite{Czarnecki:1995fw}. 
Section~\ref{twoloop} provides a detailed discussion of the NNLO contributions from $\Delta\hat{s}^2$, 
as well as the QCD 
corrections to the $\rho$ parameter, which are not captured by the running of the weak mixing angle. 
Numerical results are presented for two different renormalization schemes, where higher orders 
are parametrized in terms of powers of the fine structure constant $\alpha$ and the Fermi constant
$G_\mu$, respectively. 
The impact of hadronic uncertainties is discussed by using the framework of threshold quark 
masses introduced in Refs.~\cite{Erler:2004in,Erler:2017knj}.

Section~\ref{q2} is devoted to the analysis of residual $Q^2 \neq 0$ contributions in the self-energies. 
For this purpose a full calculation of the two-loop $\gamma\gamma$ and $\gamma Z$ self-energies has been 
performed. 
Similar to the previous section, threshold quark mass are being used to parametrize hadronic effects. 
Numerical results are presented for different values of the kinematic variables.
Our conclusions are presented in Section~\ref{conc}.


\section{NNLO corrections from the low-scale weak mixing angle}
\label{rg}
The running of the weak mixing angle in the \msbar scheme from the $Z$ scale to very low energies 
was computed in Ref.~\cite{Erler:2004in}. 
Since the RG equations (RGEs) of the vector couplings of the $Z$ boson 
and the electromagnetic coupling $\hat{\alpha}$ have a similar form, the authors could 
express the running of the weak mixing angle in terms of the running of $\hat{\alpha}$. 
To be specific,  the low energy scheme 
({\em i.e\/} 
 of  the weak mixing angle $\hat{s}(0)$) 
was defined to resum the logarithms related to $\gamma Z$ mixing. To do so, 
it was noted that such logs arise from the vector coupling of the $Z$ boson. Hence,
an analogous procedure as for the running and decoupling of $\alpha$ to the vector coupling was followed, 
allowing to properly resum such logarithms. This does not correspond to an effective field theory approach (as in the Fermi theory) below the electroweak scale, but rather we use a convenient prescription to absorb the logarithms in the  parameter $\hat{s}(0)$.
The solution of the RGE for regions between particle thresholds can be written as 
\begin{align} 
\nonumber
\hat s^2(\mu) &= \hat s^2(\mu_0) \frac{\hat\alpha(\mu)}{\hat\alpha(\mu_0)} + \lambda_1 
\left[ 1 - \frac{\hat\alpha(\mu)}{\hat\alpha(\mu_0)} \right] \\[6pt]
&+\frac{\hat\alpha(\mu)}{\pi} \left[ \frac{\lambda_2}{3} \ln\frac{\mu^2}{\mu_0^2} + \frac{3\lambda_3}{4} 
\ln \frac{\hat\alpha(\mu)}{\hat\alpha(\mu_0)} + \tilde\sigma(\mu_{0}) - \tilde\sigma(\mu) \right],
\label{eq:MASTEREQUATION}
\end{align} 
where the coefficients $\lambda_i$ are constants~\cite{Erler:2004in}
that depend on the number of particles in the theory~\cite{Erler:2004in}. 
The term $\tilde{\sigma}$ appears first at order $\hat{\alpha}^3_s$, 
and arises from OZI violating (QCD annihilation) diagrams.
In a similar way the matching conditions of $\hat{s}^2$ can also be written in terms of $\hat{\alpha}$,
\begin{equation}
   \hat{s}^2(m_i)^- = \frac{\hat\alpha(m_i)^-}{\hat\alpha(m_i)^+} 
   \hat{s}^2(m_i)^+ + \frac{Q_i T_i}{2 Q_i^2}
   \left[ 1 - \frac{\hat\alpha(m_i)^-}{\hat\alpha(m_i)^+} \right],
\label{threshold}
\end{equation}
where $Q_i$ is the electric charge of the corresponding particle, 
$T_i$ its weak isospin, and $m_i$ its mass. 
If the RGE for $\hat{\alpha}$ is solved including QCD contributions, one can obtain 
the running of the weak mixing angle from Eq.~(\ref{eq:MASTEREQUATION}) and the matching conditions. 
This resummation from $\mu=m_Z$ to $\mu=0$ is the main result of Refs.~\cite{Erler:2004in,Erler:2017knj}.
On the other hand, in Ref.~\cite{Du:2021zkj} NNLO diagrams with closed 
fermion loops were computed, but without the inclusion of QCD corrections. 
The goal is now to merge the resummed QCD corrections from Ref.~\cite{Erler:2017knj} 
and the fixed-order calculation of Ref.~\cite{Du:2021zkj}. 

To do this, we need to keep track of terms of $\mathcal{O}\left(\alpha^2\right)$ included in 
Ref.~\cite{Erler:2017knj}. 
This means we have to expand the solution to the running of the weak mixing angle up to order 
$\mathcal{O}(\alpha^2)$ with QCD effects turned off and compare with the $\mathcal{O}(\alpha^2)$ result 
for the asymmetry computed in Ref.~\cite{Du:2021zkj}.
First we obtain the expanded solution to the RGE of $\hat{\alpha}$, 
\begin{equation}
    \hat{\alpha}(\mu) = \hat{\alpha}(\mu_0) - 
    \frac{\hat{\alpha}^2(\mu_0)}{\pi}\beta_0\ln\frac{\mu^2}{\mu^2_0} + 
    \frac{\hat{\alpha}^3(\mu_0)}{\pi^2}
    \left[ \beta^2_0 \ln^2 \frac{\mu^2}{\mu^2_0} - \beta_1\ln \frac{\mu^2}{\mu^2_0}\right],
\end{equation} 
with matching condition,
\begin{equation}
   \frac{1}{\hat\alpha^+(m_f)} = \frac{1}{\hat\alpha^-(m_f)} 
   - \frac{15}{16}\, N^c_f Q_f^4\  \frac{\hat\alpha(m_f)}{\pi^2}\ ,
\label{matching}
\end{equation}
where $N^c_f$ is the color factor, and where for a single fermion with charge $Q_f$ one has 
\begin{equation} 
\beta_0 = - \frac{Q_f^2}{3}\ , \hspace{60pt}\beta_1=-\frac{Q_f^4}{4}\ .
\end{equation}
This solution can be substituted into Eq.~(\ref{eq:MASTEREQUATION}) and one obtains 
the analogous expanded RGE solution for the weak mixing angle truncated at ${\cal O}(\hat\alpha^2)$,
\begin{align} 
\label{eq:MASTEREQUATION_expanded}
\hat{s}^2(\mu) &= \hat{s}^2(\mu_0) + \frac{\hat{\alpha}(\mu_0)}{\pi}
\left( \beta_0 \lambda_1 + \frac{\lambda_2}{3} - \beta_0 \hat{s}^2(\mu_0) \right) 
\ln \frac{\mu^2}{\mu^2_0} \\[6pt]
&+ \frac{\hat{\alpha}^2(\mu_0)}{\pi^2} 
\left[ \left( \beta_1 \lambda_1 - \frac{3}{4} \beta_0 \lambda_3 - \beta_1 \hat{s}^2(\mu_0) \right)
\ln \frac{\mu^2}{\mu^2_0} - 
\left( \beta_0 \lambda_1 + \frac{\lambda_2}{3} - \beta_0 \hat{s}^2(\mu_0) \right)
\beta_0\ln^2 \frac{\mu^2}{\mu^2_0}\right].
\nonumber
\end{align}
If one wants to compute the weak mixing angle at $\mu=0$ in terms of $\hat{s}^2(m_Z)$, 
this equation should be used between particle thresholds.
For example, for $m_W < \mu < m_Z$ the QED $\beta$ function and the $\lambda_i$ constants 
include the $W$ boson. 
At $\mu=m_W$ the matching conditions for $\hat{\alpha}$ and $\hat{s}$ are used. 
Then Eq.~(\ref{eq:MASTEREQUATION_expanded}) is used again for $m_b < \mu < m_W$ but
without $W$ boson loop contributions\footnote{This just changes the values of the $\lambda_i$.}. 
This procedure is repeated until $\mu = m_e$ is reached.

With the expanded expressions for $\hat{s}^2(m_Z)$ in terms of $\hat{s}^2(0)$ up to ${\cal O}(\alpha^2)$ 
at hand, one can rewrite the semi-analytical result for $A_{\rm LR}$ from Ref.~\cite{Du:2021zkj} 
also in terms of $\hat{s}(0)$. 
The idea is to replace all occurrences of $\hat{s}(m_Z)$ by $\hat{s}(0) - [\hat{s}(0)-\hat{s}(m_Z)]$, 
where $[\hat{s}(m_z)-\hat{s}(0)]$  is the shift computed in this section, 
expanded to the required order in perturbation theory, and $\hat{s}(0)$ is the new input parameter, 
for which one can substitute the value obtained in Ref.~\cite{Erler:2017knj}.
This value includes both perturbative QCD (pQCD) corrections and non-perturbative contributions 
that enter into the RGE of the weak mixing angle, and we now briefly summarize how it was computed. 

In the perturbative regime ($\mu > 2$~GeV), solving the RGE of $\hat{s}$ is straightforward. 
Moreover, since there is no explicit $\hat{\alpha}_s$ dependence, we can also use 
Eq.~(\ref{eq:MASTEREQUATION}) for hadronic scales, and in this way the hadronic contribution to the weak 
angle is obtained from the hadronic contribution to $\hat{\alpha}$. 
For the latter, one has to rely on experimental data and dispersion 
relations~\cite{Davier:2019can,Keshavarzi:2019abf,Jegerlehner:2019lxt}.
However, different weights $\lambda_i$ enter Eq.~(\ref{eq:MASTEREQUATION}), 
because they depend on the number of active particles in the effective theory.
Thus, not only the total contribution of the three quarks to $\hat{\alpha}$ is needed, 
but also an estimate of the effective mass scales individually for the three light quarks.
This flavor separation was addressed in Ref.~\cite{Erler:2004in} by considering two limits, 
namely when the strange quark is much more massive than the up and down quarks, 
and when $\mathrm{SU}(3)$ flavor symmetry is restored. 
This issue introduced the largest source of uncertainty in the calculation. 
Later, in Ref.~\cite{Erler:2017knj} the method was refined by identifying (wherever possible) 
which channels of the $e^+e^- \rightarrow$ hadrons cross section can be associated with 
the strange quark current. 
To reduce the remaining ambiguity, lattice results~\cite{RBCUKQCD:2016clu} of the strange quark
contribution to the anomalous magnetic moment of the muon were adapted to the case at hand and included. 
The combination of flavor, data and pQCD errors gave a total uncertainty of $\pm 2\times10^{-5}$ in 
the running of $\hat{s}$ from $\mu=m_Z$ to $\mu=0$, which translates into an error of 
$\pm 8\times10^{-5}$ in the weak charge of the electron.

\section{Revisiting the one-loop result (NLO) for the asymmetry}
\label{oneloop}

For illustration we first revisit the one-loop result~\cite{Czarnecki:1995fw} for $A_{LR}$,
\begin{align} 
\nonumber
A^{\rm 1-loop}_{\rm LR} &= \frac{\rho G_\mu Q^2}{\sqrt{2}\pi\alpha}\,
\frac{1-y}{1+y^4+(1-y)^4}\bigg[1-4\kappa(0){ \hat{s}^2_Z}
\\[6pt]
\nonumber
&+ \frac{\alpha}{4\pi \hat{s}^2_Z}-\frac{3\alpha}{32\pi \hat{s}^2_Z 
\hat{c}^2_Z}(1-4{\hat{s}}^2_Z)\left(1+(1-4\hat{s}^2_Z)^2\right) \\[6pt]
&- \frac{\alpha}{4\pi}(1-4{\hat{s}}^2_Z)\left\{\frac{22}{3}\ln\frac{y 
m^2_Z}{Q^2}+\frac{85}{9}+f(y)\right\} + F_2(y,Q^2)\bigg],
\label{eq:oneloop}
\end{align}
where $\hat{s}^2_Z =
\hat{s}^2(m_Z)$ and $\hat{c}^2_Z = 1 - \hat{s}^2_Z$.
The $\rho$ parameter takes into account that the Fermi constant $G_\mu$ is obtained from 
a charged current process while polarized electron scattering is a neutral current process.
$\kappa(0)$ includes corrections from $\gamma Z$ vacuum polarization and anapole diagrams. 
The terms in the second line are from the $WW$ and $ZZ$ box contributions, respectively.
The last line arises from the $\gamma Z$ box and from photonic corrections,  
where the finite $Q^2$ effects are included in the quantity $F_2(y,Q^2)$. 
Notice that Eq.~(\ref{eq:oneloop}) is written in terms of the fine structure constant $\alpha$ 
throughout, while the original reference~\cite{Czarnecki:1995fw} employed $\hat{\alpha}(m_Z)$ 
in the $WW$ and $ZZ$ box diagrams.
We do this because the explicit two-loop calculation uses $\alpha$ in the Thomson limit as the expansion 
parameter.
The analytical expression for $\kappa(0)$ reads,
\begin{equation}
\kappa(0)\hat{s}^2_Z=\hat{s}^2_Z-\frac{\alpha}{\pi} 
\left[ \frac{1}{6} \sum_{f}\left(T_{3f}Q_f-2\hat{s}^2_ZQ^2_f\right)\ln\frac{m^2_f}{m^2_Z} -
\left( \frac{7}{4}\hat{c}_Z^2 + \frac{1}{24}\right) \ln\frac{m^2_W}{m^2_Z} + \frac{7}{18} - 
\frac{\hat{s}^2_Z}{6} \right],
\label{kappa0}
\end{equation}
while the expanded one-loop solution~\cite{Erler:2004in} to the RGE of the weak mixing angle is 
\begin{equation}
\hat{s}^2_Z-\hat{s}^2_0 = \frac{\alpha}{\pi}
\left[\frac{1}{6}\sum_{f}\left(T_{3f}Q_f-2\hat{s}^2_Z Q^2_f\right)\ln\frac{m^2_f}{m^2_Z} - 
\left(\frac{7}{4}\hat{c}^2_Z+\frac{1}{24}\right) \ln\frac{m^2_W}{m^2_Z} + \frac{1}{6} - 
\frac{\hat{s}^2_Z}{6} \right],
\label{oneloopexpanded}
\end{equation}
where $\hat{s}^2_0 =\hat{s}^2(0)$.
Hence one can see that if we insert the value $\hat{s}^2_Z = \hat{s}^2_0 + [\hat{s}^2_Z-\hat{s}^2_0]$ 
from Eq.~\eqref{oneloopexpanded} into the tree level contribution (namely $1-4\hat{s}_Z^2$) of 
the asymmetry \eqref{eq:oneloop} then there is a {\em complete} cancellation between the logarithms 
in $\kappa(0)$ and $\hat{s}^2(m_Z)-\hat{s}^2(0)$ at order $\mathcal{O}(\alpha)$. 
Effectively one can just replace 
\begin{equation}
    \kappa(0)\hat{s}^2_Z = \hat{s}^2_0 - \frac{2\alpha}{9\pi}+\mathcal{O}(\alpha^2),
\end{equation}
which corresponds to the fact that the RGE solution absorbs all logarithms originating from $\gamma Z$ 
two-point diagrams.
Since the replacement $\hat{s}_Z \rightarrow \hat{s}_0$ induces changes of order $\mathcal{O}(\alpha)$, 
such replacement applied to the terms of order $\mathcal{O}(\alpha)$ will induce effects of 
$\mathcal{O}(\alpha^2)$. 
Before studying these effects in more detail, we note that given the small number of diagrams at one-loop
and the simplicity of the result, it is not difficult to infer which scale should be used 
in certain subsets of the diagrams in order to absorb the leading higher-order contributions. 
For example, for the $\gamma Z$ mixing bubble diagrams the weak mixing angle at $\mu = 0$ should be used,
while for the $WW$ or $ZZ$ box diagrams the choice $\mu=m_Z$ provides a better approximation.
On the other hand, for $\gamma Z$ box diagrams this issue is more complicated since the loop integration encompasses all scales from 0 to $m_Z$.
If one wants to extend this kind of scale setting to two-loop order, 
one needs to identify gauge invariant subsets.
This is a challenging task which we leave for future work. 

We conclude this section by studying the numerical impact of changing $\hat{s}_Z$ to $\hat{s}_0$ 
in the $\mathcal{O}(\alpha)$ result.
It is important to note that the largest relative change will come from diagrams 
that are multiplied by a term of the form $1 - 4\hat{s}^2$, because the numerical value of $\hat{s}_Z$
is accidentally close to $1/4$. 
The $WW$ box contribution is not of this form, and there is a modest change when the replacement 
$\hat{s}_Z \rightarrow \hat{s}_0$ is applied ($\Delta Q_W^e \approx 0.08\times10^{-3}$). 
The $ZZ$ box contribution to the weak charge is small to begin with, and even though the replacement 
produces a large relative change, it results in only a small change 
of $\approx 0.04\times10^{-3}$ in $Q_W^e$. 
The largest numerical impact of the replacement is due to $\gamma Z$ box diagrams.
This was first discussed in Ref.~\cite{Czarnecki:1995fw}, and due to the ambiguity in scale setting 
mentioned in the previous paragraph, the authors assigned half of the difference between 
using $\hat{s}_0$ and $\hat{s}_Z$ in the $\gamma Z$ box as a perturbative uncertainty,
amounting to about $10^{-3}$ in $Q_W^e$.

\section{Two-loop results (NNLO)}
\label{twoloop}
In this section, numerical results are presented in two renormalization schemes, 
which differ in the way the electroweak coupling is renormalized,
\begin{align}
&\alpha \text{ scheme}:
& \Delta Q_W^e &= \alpha \, \Delta Q_{W(1)}^{e,\alpha} + \alpha^2 \, \Delta Q_{W(2)}^{e,\alpha}, 
\label{eq:alp} \\
&G_\mu \text{ scheme}:
& \Delta Q_W^e &= G_\mu \, \Delta Q_{W(1)}^{e,G} + G_\mu^2 \, \Delta Q_{W(2)}^{e,G}. 
\label{eq:Gmu}
\end{align}
In Eq.~\eqref{eq:alp} the electromagnetic coupling is renormalized in the Thomson limit, 
which introduces a dependence of the final result on the shift $\Delta\alpha$ that accounts for the 
effective running of the fine structure constant between the scale $\mu=0$ and $\mu=m_Z$. 
The weak coupling in this scheme is defined as $g = e/\hat{s}$.

The translation to the $G_\mu$ scheme is accomplished by using the relation
\begin{align}
\frac{G_\mu}{\sqrt{2}} = \frac{\pi\alpha}{2\hat s_Z^2\hat c_Z^2 m_Z^2}(1+\Delta r),
\end{align}
where $\Delta r$ accounts for radiative corrections. 
The electroweak two-loop corrections to $\Delta r$ have been taken from
Refs.~\cite{Freitas:2000gg,Freitas:2002ja} (see also Ref.~\cite{Awramik:2003ee}).

In the following, we use the notation $\Delta Q_{W(L,n_f)}^{e,X}$ with $X=\alpha,G_\mu$
to further distinguish the $L$-loop corrections by the number $n_f$ of closed fermion loops,
\begin{align}
\Delta Q_{W(1)}^{e,X} &= \Delta Q_{W(1,1)}^{e,X} + \Delta Q_{W(1,0)}^{e,X}\, , 
&\Delta Q_{W(2)}^{e,X} &= \Delta Q_{W(2,2)}^{e,X} + \Delta Q_{W(2,1)}^{e,X}\, .
\end{align}
The two-loop corrections without closed fermion loops, $\Delta Q_{W(2,0)}^{e,X}$,
are not known at this time, but they have been estimated to be of ${\cal O}(10^{-4})$~\cite{Du:2021zkj}. 

The NNLO result of Ref.~\cite{Du:2021zkj} is given in terms of $\hat{s}(m_Z)$. 
There the \msbar counterterm $\delta\hat{s}^2_Z$ at $\mu=m_Z$ is calculated in the full six-flavor SM.
To rewrite our expressions in terms of $\hat{s}^2_0$ we define
$\delta \hat{s}^2_0\equiv \delta \hat{s}^2_Z-\Delta\hat{s}^2$ where 
$\Delta\hat{s}^2=\hat{s}^2_0-\hat{s}^2_Z$. 

For our numerical results we use the values,
\begin{align}
m_Z &= 91.1876\text{ GeV}, & m_H &= 125.1 \text{ GeV}, \nonumber \\
m_\tau &= 1.777 \text{ GeV}, & m_t &= 173.0 \text{ GeV}, & m_b &= 3.99 \text{ GeV}, \nonumber \\
m_\mu &= 105.7 \text{ MeV}, & m_c &= 1.185 \text{ GeV}, & m_s &= 342 \text{ MeV}, \nonumber \\
m_e &= 0.511 \text{ MeV}, & m_{u,d} &= 246 \text{ MeV}, \nonumber  \\
\alpha^{-1} &= 137.036, & \Delta\alpha &= 0.02761_\text{had}+0.0314976_\text{lep},  \nonumber \\
\hat{s}^2(m_Z) &= 0.2314, & \hat{s}^2(0) &= 0.23861, \nonumber \\
& & G_\mu &= 1.1663787 \times 10^{-5} \text{ GeV}^{-2}.
\label{eq:input1}
\end{align}
The values and uncertainties for $\hat{s}(0)$, the light quark masses $m_q$, $q \neq t$, and the hadronic
contribution to $\Delta\alpha$, are taken from the RG analysis of Ref.~\cite{Erler:2017knj}.  
The leptonic contribution to $\Delta\alpha$ has been computed perturbatively to 4-loop accuracy 
in Ref.~\cite{Sturm:2013uka}.  Furthermore, considering the MOLLER experiment with an electron beam 
energy of $E_{\rm beam}=11$~GeV, the center-of-mass energy is given by 
$s=2m_eE_{\rm beam}=0.011$~GeV$^2$. 
With these inputs, we obtain the numerical results in Table~\ref{tab:resAv}.

The result in the $\hat{s}(m_Z)$--$\alpha$ scheme corresponds to column 2 in Table~\ref{tab:resAv}. 
The numbers shown there are identical to the ones reported in Ref.~\cite{Du:2021zkj} where the weak-scale
mixing angle $\hat{s}(m_Z)$ was used without the inclusion of pQCD. As for the non-perturbative 
(hadronic) effects, some of these are included in the threshold masses, 
but others are missing as we explain in what follows.

The parametrization~\cite{Erler:2017knj} to incorporate the non-perturbative light quark contributions 
to $\Delta\hat{\alpha}$ includes --- in addition to the light quark threshold masses --- 
a second parameter $K_q$, in terms of which the contribution of a light quark has the form
\begin{equation}
\Delta\hat{\alpha}\sim \frac{\alpha }{\pi}Q_q^2K_q\ln\frac{\mu^2}{m^2_q}\ .
\label{eq:kq}
\end{equation}
At first sight, the parameter $K_q$ may seem redundant, as it is the combination \eqref{eq:kq} 
as a whole that is constrained by $e^+ e^-\rightarrow\,\mathrm{hadrons}$ data.
However, there is a monotony constraint on the $K_q$ since $K_1 > K_2$ for two quarks with masses
$m_1 < m_2$ (for asymptotically large quark masses $K_q \to 1$), and in addition $K_c$ and $K_b$ can be 
computed in pQCD.
This additional information reduces the uncertainty but introduces a large correlation between
$m_q$ and $K_q$.
The second column in Table~\ref{tab:resAv} does not contain the $K_q$ effects since $K_q = 1$ was 
assumed. 

\begin{table}[t]
\centering
\renewcommand{\arraystretch}{1.3}
\begin{tabular}{|c|c|cc|}
\hline
 & $\hat{s}(m_Z)$--$\alpha$ scheme$^*$~\cite{Du:2021zkj}
 & $\hat{s}(0)$--$\alpha$ scheme 
 & $\hat{s}(0)$--$G_\mu$ scheme \\[-.5ex]
 & \multicolumn{1}{c|}{$(X{=}\alpha)$} 
 & \multicolumn{1}{c}{$(X{=}\alpha)$} 
 & \multicolumn{1}{c|}{$(X{=}G_\mu)$} \\
\hline
$1-4\hat{s}^2$ & $\phantom{+}74.40$ & $\phantom{+}45.56$ & $\phantom{+}45.56$\\
\hline
$X\,\Delta Q_{W(1,1)}^{e,X}$ & $-29.04$ & $+\phantom{0}0.39$ & $+\phantom{0}0.43$ \\
$X\,\Delta Q_{W(1,0)}^{e,X}$ & $+\phantom{0}3.06$ & $+\phantom{0}0.77$ & $+\phantom{0}0.84$ \\
\hline
$X^2\,\Delta Q_{W(2,2)}^{e,X}$ & $-\phantom{0}0.18$ & $+\phantom{0}0.07$ & $+\phantom{0}0.05$ \\
$X^2\,\Delta Q_{W(2,1)}^{e,X}$ & $+\phantom{0}1.18$ & $-\phantom{0}1.15$ & $-\phantom{0}1.30$ \\
\hline
$X\,\Delta Q_{W, \Delta\rho}^{e,X}$ & \quad --- & $-\phantom{0}0.05$ & $-\phantom{0}0.06$ \\
\hline
Sum & $\phantom{+}49.42$ & $\phantom{+}45.60$ & $\phantom{+}45.52$ \\
\hline
\multicolumn{1}{c}{} & \multicolumn{1}{c}{$^*$no QCD corrections}
\end{tabular}
\caption{Corrections (in units of $10^{-3}$) at different orders contributing to the electron's weak 
charge in polarized M\o ller scattering in three different input schemes for the SM input parameters in 
Eq.~\eqref{eq:input1} and the kinematic parameters $s = 0.011$~GeV$^2$ and $y=0.4$.
In our notation, $\Delta Q_{W(L,n_f)}$ is the contribution at $L$-loop order from diagrams with $n_f$ 
closed fermion loops.} 
\label{tab:resAv}
\end{table}
The third and fourth columns contain the results when using the low-scale mixing angle $\hat{s}(0)$ 
as input in the two renormalization schemes defined in Eqs.~\eqref{eq:alp} and \eqref{eq:Gmu}. 
We now discuss the table row by row. 

\begin{itemize}
\item \underline{$1-4\hat{s}^2$:} 
This is the tree level contribution. We can see a large difference 
between column two and the low energy schemes (columns three and four).  
This is because the weak mixing angle at zero absorbs all large logarithms $\sim \ln(m_f^2/m_Z^2)$. 
The error associated to this contribution comes from the error in the weak mixing angle at zero 
momentum~\cite{Erler:2017knj}, and translates to an error of $\pm 0.08\times10^{-3}$ in the weak charge. 

\item \underline{$X\,\Delta Q_{W(1,1)}^{e,X}$:} 
By comparing column two  of Table~\ref{tab:resAv}, with the low energy schemes, one can immediately see 
that the size of the NLO corrections is reduced by more than an order of magnitude in the low energy 
schemes, which is mostly due to the absence of large logarithms $\sim \ln(m_f^2/m_Z^2)$, 
\textit{i.e.}, these logarithms are already absorbed in the tree-level result.
The error induced in the weak charge by the error on the input parameter $\hat{s}^2(0)$ is negligible for 
these type of diagrams.

\item \underline{$X\,\Delta Q_{W(1,0)}^{e,X}$:} 
The difference relative to column two comes mainly from the $\gamma Z$ box. 
There is also a reduction due to the fact that $\hat{s}^2(0)$ contains the W-boson contribution 
to the $\gamma Z$ bubble. 
The error induced in the weak charge by the error on the input parameter $\hat{s}^2(0)$ is negligible for 
these type of diagrams.

\item \underline{$\Delta Q^{e,X}_{W(2,2)}$:} 
A reduction also emerges for the contributions with two closed fermion loops. 
Part of this is due to the resummation of the fermionic logarithms in $\hat{s}(0)$. 
In the $\hat{s}(0)$--$\alpha$ scheme, the logarithmic dependence on the fermion masses is not completely 
cancelled since the weak mixing angle counterterm also appears in subloop renormalization contributions, 
without any connection to the t-/u-channel $\gamma Z$ self-energy. 
It turns out that these additional appearances of the weak mixing angle counterterm are cancelled 
in the $\hat{s}(0)$--$G_\mu$ scheme, so that $\Delta Q^{e,X}_{W(2,2)}$ does not have any dependence on 
the light fermion masses in this scheme, \textit{i.e.}, all the logarithms drop out. 
This is a consequence of the similarity of the charged-current Fermi interaction and the neutral-current 
parity-violating contribution to M\o ller scattering, which are both weak four-fermion
processes\footnote{Moreover, the result for $\Delta Q^{e,X}_{W(2,2)}$ in the $\hat{s}(0)$--$G_\mu$ scheme
is not only independent of the quark mass logarithms stemming from the $\gamma Z$ self-energy, 
but in addition is independent of the shift in the fine structure constant, $\Delta\alpha$.}. 
Hence, we have the picture that in the $\hat{s}^2(0)$--$G_\mu$ scheme the dependence on the light quark masses 
is completely removed, while in the $\hat{s}^2(0)$--$\alpha$ scheme a small dependence on the quark 
masses remains. 
To estimate the remaining hadronic uncertainty (that is not already taken into account in $\hat{s}^2(0)$)
we vary these masses within the ranges given in Ref.~\cite{Erler:2017knj}, 
translating into a negligible error in the weak charge in the $\hat{s}^2(0)$--$\alpha$ scheme.

\item \underline{$\Delta Q^{e,X}_{W(2,1)}$:} 
For the contributions with only one closed fermion loop, we did not find any significant reductions in 
the corrections or the dependence on the fermion masses (cancellation of logarithms). 
While this may seem surprising at first glance, one must keep in mind that the $\Delta Q^{e,X}_{W(2,1)}$ 
corrections have a much more complicated structure than the contributions with two closed fermion loops. 
For instance, they contain two-loop vertex and box diagrams, exemplified in Figure~\ref{fig:subl}, 
that depend on the fermion masses in a non-trivial way. 
These diagrams have been computed numerically in Ref.~\cite{Du:2021zkj} 
using a dispersion relation for the fermionic subloops. 
The integration region spans all values of $|k^2|$ from 0 to $\infty$, where $k$ is the momentum
flowing through the fermion subloops, 
while $\hat{s}(0)$ only absorbs the fermion mass dependence at $k^2=0$.
Therefore one should not expect any significant cancellations in the $\hat{s}(0)$ schemes. 
Indeed, the threshold masses obtained in Refs.~\cite{Erler:2004in,Erler:2017knj} were constructed to 
quantify the total hadronic contribution to the running of the weak mixing angle between $m_c$ and zero. 
On the other hand, within the hadronic region this parametrization may lose its 
justification. To account for this additional theoretical uncertainty, we take a conservative approach and assign 
a factor of two error in the masses\footnote{If one took the nominal values from Ref.~\cite{Du:2021zkj} instead, this uncertainty would be $\pm 0.03\times10^{-3}$.}, $m_q{}^{+1.0 m_q}_{-0.5m_q}$,
translating into an hadronic error in $\Delta Q^{e,X}_{W(2,1)}$ of less than $\pm 0.06\times10^{-3}$. 
 A more refined possibility is to use the vacuum polarization functions obtained from
$e^+e^-$ data using dispersive techniques or lattice results \cite{Ce:2022eix}, as described for example in Ref.~\cite{Proceedings:2019vxr}. In this approach it is possible, with certain
theory assumptions about flavor separation, to obtain $\Pi_
{\gamma \gamma}(s)$ and $\Pi_{\gamma Z}(s)$ in the hadronic region. These
can then be inserted in the two-loop diagrams and numerically integrated. There are recent developments in the calculation of such integrals in the context of $\mu e$ scattering in both the timelike \cite{Fael:2019nsf} and the spacelike regions \cite{Fael:2018dmz}. 
One should then understand our uncertainty as a conservative estimate of the size of the hadronic effects for these type of diagrams\footnote{The reader might be worried that perhaps the full vacuum polarisation function is also needed in the $\Delta Q^{e,X}_{W(1,1)}$  and $\Delta Q^{e,X}_{W(2,2)}$ contributions discussed previously, given the finite $q^2$ of MOLLER.  That is, corrections which go as $q^2/m^2_q$, or more properly $q^2/m^2_\pi$.  Nevertheless, as we explain in the next section, given the kinematic values at MOLLER we estimated such finite $q^2$ terms to be rather small, so the detailed form of the vacuum polarisation function in such diagrams is a subleading effect and well under control, and the use of $\chi \mathrm{PT}$, threshold masses or a full integral expression of $\Pi(q^2)$ do not significantly change the result. This is in contrast with the process calculated in 
Ref.~\cite{Fael:2019nsf} ($\mu e$ scattering) where the momentum transfer is in the range 
$-0.143\, \mathrm{GeV}^2\, <\, t\, <\, 0$.}.
\begin{figure}[t]
\centering
\epsfig{figure=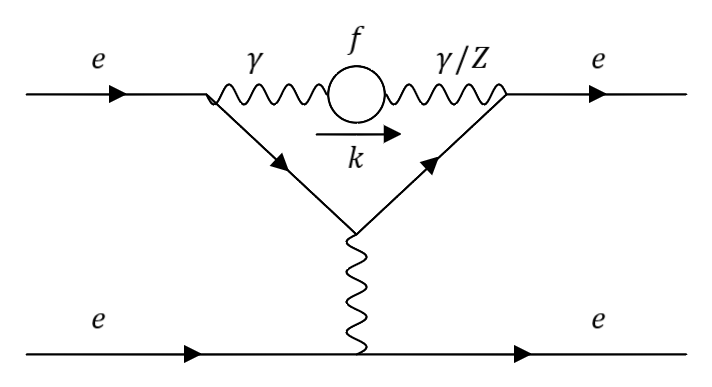, width=6.4cm}
\hspace{1cm}
\epsfig{figure=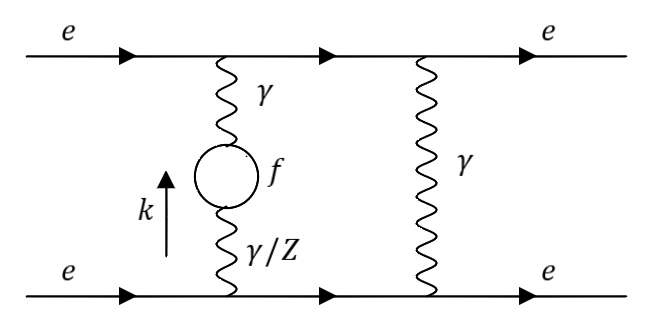, width=6.4cm}
\caption{Two-loop diagrams contributing to M\o ller scattering with a
non-trivial dependence on light fermion masses.}
\label{fig:subl}
\end{figure}

\item \underline{QCD corrections:} 
The bulk of the higher-order QCD corrections are captured by the RG running of $\hat{s}(\mu)$, 
which sums up powers of large fermionic logarithms to all orders. 
The evaluation of $\hat{s}(0)$ in Ref.~\cite{Erler:2017knj} includes RG effects up to ${\cal O}(\as^4)$. 
A separate source of QCD corrections enters through the $\rho$ parameter, which describes 
$m_t^2$--enhanced contributions due to custodial symmetry breaking. 
It contributes
\begin{align}
\Delta Q^e_W = (1-4\sin^2\theta_W)\Delta\rho + \cdots \ ,
\end{align}
where the dots indicate the remaining radiative corrections. 
The leading contribution is given by
\begin{align}
\Delta\rho_{(\alpha)} = \frac{3\alpha m_t^2}{16\pi \hat{s}^2 m_W^2}~~~(\alpha \text{ scheme}), &&
\Delta\rho_{(G_\mu)} = \frac{3G_\mu m_t^2}{8\sqrt{2}\pi^2}~~~(G_\mu \text{ scheme}),
\end{align}
and is included in the NLO correction $\Delta Q^{e,X}_{W(1,1)}$. 
QCD corrections from higher orders are given 
by~\cite{Djouadi:1987gn,Djouadi:1987di,Kniehl:1989yc,Avdeev:1994db,Chetyrkin:1995ix,Chetyrkin:2006bj},
\begin{align}
\Delta\rho_{(X\as^n)} &= \Delta\rho_{(X)}\biggl[
 - \frac{4}{3}\Bigl( \frac{1}{2}+\frac{\pi^2}{6}\Bigr)\frac{\as}{\pi}
 - 14.594\Bigl(\frac{\as}{\pi}\Bigr)^2 
 - 93.15\Bigl(\frac{\as}{\pi}\Bigr)^3 
 + {\cal O}(\as^4) \biggr], 
\end{align}
where $X=\alpha,G_\mu$. 
We denote these higher-order effects as
\begin{align}
X\,\Delta Q_{W, \Delta\rho}^{e,X}\equiv (1-4\hat{s}^2)\Delta\rho_{(X\as^n)}\ .
\end{align}
With $\as(m_Z) = 0.1182$ one obtains the corrections listed in Table~\ref{tab:resAv}. 
The propagation of an error $\delta \as(m_Z)=0.0016$ in these terms translates to $10^{-6}$ for 
the weak charge, which is negligible.  

\item \underline{Missing contributions $\Delta Q^{e,X}_{W(2,0)}$:}
The uncertainty in the weak charge from the missing purely bosonic NNLO corrections was 
estimated in Ref.~\cite{Du:2021zkj} to about $\pm1.3 \times 10^{-4}$.

\item \underline{Higher order electroweak contributions:}  
As can be seen from the last row of the table, the total prediction for the electron weak charge differs 
significantly between the $\hat{s}(m_Z)$ scheme and the two $\hat{s}(0)$ schemes 
($49.4 \times 10^{-3}$ {\em vs.\/} $45.6 \times 10^{-3}$). 
This difference can be attributed mainly to QCD corrections, which are missing in the former but included
in the latter. 
The difference between the $\hat{s}(0)$--$\alpha$ and $\hat{s}(0)$--$G_\mu$ schemes could be regarded as 
an estimate of the theory error from missing higher orders (NNNLO and beyond). 
Its magnitude of slightly less than $10^{-4}$ is comparable to the theory error estimate 
for the $\Delta Q^{e,X}_{W(2,0)}$ contribution. Another way to estimate the higher-order electroweak 
uncertainty is to take the difference between the $\hat{s}(m_Z)$ scheme and any of the $\hat{s}(0)$ 
schemes. 
But in order to do this one has to include the same QCD contributions in both schemes. 
Therefore, we first re-computed column two of Table~\ref{tab:resAv} including the non-perturbative 
effects\footnote{We did this by absorbing these effects into the phenomenological masses which lowers 
their values.} contained in $K_q$. 
This would include all non-perturbative effects, but would miss the resummation of the logarithms and 
the dependence on pQCD so that it can be compared with a re-computed column three ($\hat{s}(0)$ scheme) 
with pQCD turned off.
The difference between these re-computed weak charges in the two schemes ($\hat{s}^2(0)$--$\alpha$ and 
$\hat{s}^2(m_Z)$--$\alpha$) should then be due to electroweak effects. 
We found that this difference is $4.5\times10^{-4}$, which is much smaller than the difference between 
columns two and three in Table~\ref{tab:resAv}, demonstrating that the latter is mainly due to the
missing QCD effects in column two\footnote{This also served as a double-check of the 
implementation of our schemes.}. 
The remaining difference may be attributed to the scheme choice. 
Bearing in mind that this difference is obtained in a similar way to how Ref.~\cite{Czarnecki:1995fw} 
estimated the error from the $\gamma Z$ box, we may take the interval spanned by the results 
in the two schemes as a conservative estimate of the higher-order perturbative error on the weak charge. 
Adding back the  pQCD contributions, this implies that we have the error interval 
$[45.60,45.60+0.45]\times 10^{-3}$ or $(45.83\pm 0.23)\times 10^{-3}$, 
where the lower bound corresponds to the $\hat{s}^2(0)$--$\alpha$ result.
An alternative error estimation is obtained by studying the shifts induced by the change of the weak 
mixing angle from $\hat{s}(m_Z)$ to $\hat{s}(0)$ in the $\Delta Q_{W(2,1)}$ terms, leaving everything 
else fixed, which results in a shift of similar size ($2\times10^{-4}$) for the weak charge.  
\end{itemize}
In summary, we find for the weak charge of the electron,
\begin{align}\nonumber
    Q^e_W &= (45.83 \pm 0.08_{\, \hat s(0)} 
    \pm0.06_{\Delta Q^{e,X}_{W(2,1)}(\mathrm{\, had})}
    \pm0.13_{\Delta Q^{e,X}_{W(2,0)}(\mathrm{\, missing})}
    \pm0.23_{\mathrm{\, scheme}}) \times 10^{-3} \\
    &= \left(45.83\pm0.28_\mathrm{\, theory}\right)\times10^{-3},
    \label{eq:WeakCharge}
\end{align}  
where in the last line we added all errors in quadrature.
Comparing with the experimental precision expected at MOLLER, $\delta Q_W^e=1.1\times10^{-3}$,
we see that the theoretical error is under control. In this result, the correlations between the light quark masses  are properly included in the error of $\hat{s}^2_0$ computed in Ref.~\cite{Erler:2017knj}. For the calculation of the box and vertex diagrams, the errors on $m_{u,d}$ and $m_s$ are assumed to be fully anticorrelated, which is a simplifying but conservative assumption.
We emphasize that the numerical difference between our result and Ref.~\cite{Du:2021zkj} is due to the inclusion of QCD effects, both perturbative and non-perturbative contributions parametrized by the $K_q$ in Ref.~\cite{Erler:2017knj}. To understand the impact of the scheme choice, the QCD corrections must be treated on equal footing in both schemes. This comparison has been carried out in this paper, and the difference between both schemes is taken to define the scheme error in Eq.~(\ref{eq:WeakCharge}). Furthermore, it is important to remark that taking half the difference between the $\hat{s}^2(0)$ and 
$\hat{s}^2(m_Z)$ schemes likely overestimates the perturbative error since we expect the exact 
(all orders) result to be closer to the low-scale schemes. 
On top of that, we believe that the estimation of the electroweak perturbative error can be better 
understood and further reduced through a more careful resummation of dominant diagrams. 
Such an analysis requires the study of gauge invariant diagram subsets
which is a complicated task and left for future work.

\section{Finite momentum transfer effects}
\label{q2}

In the calculation of Ref.~\cite{Du:2021zkj}, the momentum transfer through the
t- and u-channel propagators was approximated to be zero, $Q^2 \to 0$. 
At the one-loop level, it was found that the shift in the transverse self-energies, 
$\Pi_{\rm T}^{\gamma\gamma}(-Q^2)-\Pi_{\rm T}^{\gamma\gamma}(0)$
and $\Pi_{\rm T}^{\gamma Z}(-Q^2)-\Pi_{\rm T}^{\gamma Z}(0)$, is very small for the
kinematic parameters of the E158 and MOLLER experiments~\cite{Czarnecki:1995fw}.
However, it is worth verifying that this also holds at two loops. 
This is clearly the case for diagrams where all particles in the loop have large
masses, $m_i^2 \gg Q^2$, since any momentum-dependent term scales like
$Q^2/m_i^2$ in these contributions. But for diagrams with light fermions
($e,\mu,u,d,s$) in the loop it is less obvious that $Q^2 \to 0$ is a good
approximation.

To investigate this question, we have computed the relevant $\gamma\gamma$ and
$\gamma Z$ one- and two-loop self-energies for $Q^2 \neq 0$. The one-loop
self-energies allow us to reproduce and verify the results of
Ref.~\cite{Czarnecki:1995fw}, while the two-loop self-energies will be used to
study the quality of the $Q^2 \to 0$ approximation used in
Ref.~\cite{Du:2021zkj}. At two-loop order, or NNLO, one also needs to include
one-particle reducible diagrams with a one-loop self-energy and a
one-loop vertex correction, as well as the interference of two one-loop amplitudes.
We restrict ourselves to NNLO contributions with
at least one closed fermion loop, since this is the order of corrections
considered in Ref.~\cite{Du:2021zkj}. Moreover, the self-energy diagrams without
fermions do not contain any particles with masses comparable to or below $Q^2$.

The package {\sc FeynArts 3}~\cite{Hahn:2000kx} has been used for generating the
amplitudes for the one- and two-loop self-energies. The Lorentz and Dirac
algebra has been performend with an in-house code, implemented in {\sc
Mathematica}. This code also performs a reduction to a set of master integrals,
based on the technique of Ref.~\cite{Weiglein:1993hd}. The master integrals can
be evaluated numerically with TVID~2~\cite{Bauberger:2019heh}, which uses the
one-dimensional integral representations developed in
Ref.~\cite{Bauberger:1994by,Bauberger:1994hx}.

As in the previous section, the hadronic self-energy contributions are described
by computing quark loops and using the threshold quark masses from
Ref.~\cite{Erler:2017knj} that have been derived from a renormalization-group
analysis. 
As shown in Eq.~(\ref{eq:oneloop}), the impact of the $Q^2$-dependence of the self-energies on the
asymmetry $A_{\rm LR}$ can be written as~\cite{Czarnecki:1995fw}
\begin{align}
A_{\rm LR} = \frac{G_\mu Q^2}{\sqrt{2}\pi
\alpha}\,\frac{1-y}{1+y^4+(1-y)^4}[1-4\sin^2\theta_W + F_2(Q^2,y) + ...],
\end{align}
where $y = Q^2/s$ and the dots denote all other higher-order corrections. By
construction, $F_2(0,y)=0$.

\begin{table}[tb]
\centering
\renewcommand{\arraystretch}{1.2}
\begin{tabular}{|c|c|cc|}
\hline
$y$ & $F_2^{(1,1)}\;[10^{-5}]$ & $F_2^{(2,2)}\;[10^{-5}]$ & $F_2^{(2,1)}\;[10^{-5}]$ \\
\hline
0.25 (0.75) & 5.01 & $-$0.54 & \phantom{$-$}0.00 \\
0.30 (0.70) & 4.11 & $-$0.25 & $-$0.18 \\
0.35 (0.65) & 3.39 & $-$0.02 & $-$0.29 \\
0.40 (0.60) & 2.86 & \phantom{$-$}0.15 & $-$0.36 \\
0.45 (0.55)  & 2.55 & \phantom{$-$}0.25 & $-$0.40 \\
0.50  & 2.44 & \phantom{$-$}0.29 & $-$0.41 \\
\hline
\end{tabular}
\caption{Numerical results for $F_2(Q^2,y)$, which captures the $Q^2$-dependence
of the photon and photon--Z self-energies. $F_2^{(L,n_f)}$ denotes corrections
with $L$ loops and $n_f$ closed fermion loops. Results are shown as a function
of $y=Q^2/s$, for $s=0.011$~GeV$^2$ and SM input parameters in Eq.~\eqref{eq:input1}.}
\label{tab:F2}
\end{table}
With the input parameters in Eq.~\eqref{eq:input1}, as well as $s=0.011$~GeV$^2$,
the numerical results listed in Table~\ref{tab:F2} are obtained. The 1-loop
contribution for $y=0.5$ agrees well with the analysis of
Ref.~\cite{Czarnecki:1995fw}, which found $F_2(y = 0.5) \approx 2 \times
10^{-5}$. For all experimentally relevant values of $y$, the NLO contributions
to $F_2$ stay well below $10^{-4}$ and thus are irrelevant for practical
purposes. The NNLO contributions can be divided into terms with two and one
closed fermion loop. Both of these, as well as the sum of the NNLO effects, are
about one order of magnitude smaller than the NLO contributions. This confirms
that the $Q^2 \to 0$ approximation used in Ref.~\cite{Du:2021zkj} is accurate
and robust at NLO and NNLO.


\section{Conclusions}
\label{conc}

The left-right polarization asymmetry in M\o ller scattering is a sensitive probe of parity violation in 
the SM and from new physics.
Recently, the SM electroweak two-loop corrections from contributions with closed fermion loops to
this observable were computed in Ref.~\cite{Du:2021zkj}. 
For phenomenlogical applications, this result needs to be combined with resummed QCD and hadronic 
effects, which can be incorporated through the renormalization group analysis of the \msbar\ weak mixing 
angle $\hat{s}(\mu)$ at low scales $\mu \approx 0$~\cite{Erler:2004in,Erler:2017knj}. 
Two new schemes are introduced, labeled $\hat{s}(0)$--$\alpha$ and $\hat{s}(0)$--$G_\mu$, respectively. 
Both use the low-energy \msbar\ weak mixing angle as input, but the former scheme uses $\alpha$ for 
the power counting of the electroweak perturbative expansion, whereas the latter uses $G_\mu$. 

In addition to the perturbative and non-perturbative QCD effects from the running of $\hat{s}(\mu)$, we 
also include perturbative QCD corrections to the $\rho$ parameter. 
Finally, we carry out a careful analysis of the dependence of the scattering rate on the momentum 
transfer squared $Q^2$, which we find to be numerically negligible. The SM prediction for the left-right 
asymmetry, including higher-order effects, can be expressed in terms of the weak charge $Q_W^e$ 
of the electron. 
In the $\hat{s}(0)$--$\alpha$ scheme we obtain $Q_W^e=(45.83\pm0.28)\times10^{-3}$, 
where the dominant error of $\pm 0.23\times10^{-3}$ stems from the purely electroweak difference between the $\hat{s}(m_Z)$-$\alpha$ and 
$\hat{s}(0)$-$\alpha$ schemes. 
The result is consistent with the one-loop calculation of Ref.~\cite{Czarnecki:1995fw} and implies 
a reduction of the uncertainty by almost an order of magnitude. Additional relevant sources of 
uncertainty stem from the currently unknown bosonic two-loop corrections 
(estimated as $\delta Q^e_W = \pm 0.13 \times 10^{-3}$~\cite{Du:2021zkj}) 
and from the running of $\hat{s}(\mu)$  including non-perturbative effects
(estimated to amount to $\delta Q^e_W = \pm 0.08 \times 10^{-3}$~\cite{Erler:2017knj}).
When compared to the expected precision $\delta Q_W^e=1.1\times10^{-3}$ 
of the planned MOLLER experiment~\cite{Benesch:2014bas}, the overall uncertainty of the SM prediction 
turns out to be insignificant. 
Furthermore, this theoretical error is rather conservative, since the low-energy scale of the process 
suggests that the $\hat{s}(0)$ scheme is the more adequate one to use in the tree-level expression, 
so that considering half the scheme difference may overestimate the uncertainty. 
Future work on enhanced three-loop effects and the purely bosonic two-loop corrections can be expected
to reduce the theory error further. 


\section*{Acknowledgments}
This work has been supported in part by the National Science Foundation under grants no.\ PHY-1820760 and PHY-2112829.


\bibliographystyle{JHEP}
\bibliography{moller}

\end{document}